\documentclass[superscriptaddress,twocolumn, 10pt, nofootinbib]{revtex4}
\usepackage{amsmath,lscape,sidecap,epsfig}
\usepackage{graphicx,color,xcolor}
\usepackage{amssymb}	

\usepackage{color,xcolor}
\usepackage{lipsum}
\usepackage{multirow}
\usepackage{hyperref}

\DeclareFontFamily{U}{dutchcal}{\skewchar\font=45 }
\DeclareFontShape{U}{dutchcal}{m}{n}{<-> s*[1.0] dutchcal-r}{}
\DeclareFontShape{U}{dutchcal}{b}{n}{<-> s*[1.0] dutchcal-b}{}
\DeclareMathAlphabet{\mathlcal}{U}{dutchcal}{m}{n}
\SetMathAlphabet{\mathlcal}{bold}{U}{dutchcal}{b}{n}

\begin{document}

\title{Exploring the effects of Delta Baryons in magnetars}

\author{K. D. Marquez}
\email[]{marquezkauan@gmail.com}
\affiliation{Depto de F\'{\i}sica - CFM - Universidade Federal de Santa
Catarina  Florian\'opolis - SC - CP. 476 - CEP 88.040 - 900 - Brazil}

\author{M. R. Pelicer}
\affiliation{Depto de F\'{\i}sica - CFM - Universidade Federal de Santa
Catarina  Florian\'opolis - SC - CP. 476 - CEP 88.040 - 900 - Brazil}

\author{S. Ghosh}
\affiliation{Inter-University Centre for Astronomy and Astrophysics, Post Bag 4, Ganeshkhind, Pune
University Campus, Pune, 411007, India}

\author{J. Peterson}
\affiliation{Department of Physics, Kent State University, Kent, OH 44243, USA}

\author{D. Chatterjee}
\affiliation{Inter-University Centre for Astronomy and Astrophysics, Post Bag 4, Ganeshkhind, Pune
University Campus, Pune, 411007, India}

\author{V. Dexheimer}
\affiliation{Department of Physics, Kent State University, Kent, OH 44243, USA}

\author{D. P. Menezes}
\affiliation{Depto de F\'{\i}sica - CFM - Universidade Federal de Santa
Catarina  Florian\'opolis - SC - CP. 476 - CEP 88.040 - 900 - Brazil}

\begin{abstract}
{Strong magnetic fields can modify the microscopic composition of matter with consequences on stellar macroscopic properties. Within this context, we study, for the first time, the possibility of the appearance of spin-3/2 $\Delta$ baryons in magnetars. We make use of two different relativistic models for the equation of state  of dense matter under the influence of strong magnetic fields considering the effects of Landau levels and the anomalous magnetic moment (AMM) proportional to the spin of 
all baryons and leptons. In particular, we analyze the effects of the AMM of $\Delta$ baryons in dense matter for the first time. {We also obtain global properties corresponding to the EoS models numerically and study the corresponding role of the $\Delta$ baryons.} We find that they are favored over hyperons, which causes an increase in isopin asymmetry and a decrease in spin polarization. We also find that, contrary to what generally occurs when new degrees of freedom are introduced, the $\Delta$s do not make the EoS significantly softer and magnetars less massive. Finally, the magnetic field distribution inside a given star is not affected by the presence of $\Delta$s.}
\end{abstract}


\maketitle

\section{Introduction}

Magnetars are a class of compact objects that possess the largest stable magnetic fields observed in nature, with surface magnitudes inferred for the poloidal component in the range of $10^{11}-10^{15}$ G at the surface \citep{haensel2006neutron} and values more than one  order of magnitude larger in the interior \citep{PhysRevLett.112.171102,10.1093/mnras/sty1706}. 
Although the strength of the magnetic field in the central region of these stars remains unknown, they could reach $\sim10^{18}$ G according to the scalar virial theorem \citep{1991ApJ...383..745L,1993A&A...278..421B}, and simultaneous solutions of Einstein and Maxwells equations for poloidal \citep{1995A&A...301..757B,Cardall:2000bs} and also toroidal configurations \citep{10.1093/mnras/stx1176,Tsokaros:2021pkh}.
Such extreme conditions certainly play a considerable role when determining the internal composition and structure of magnetars.

The starting point for determining the macroscopic structure of compact stars is the assumption of a specific microscopic model, which leads to the calculation of an equation of state (EoS) for dense matter. 
The EoS encodes the particle population of baryons and leptons and how they interact through the strong interactions, constrained by equilibrium conditions, such as $\beta$-stability and charge neutrality. 
The extremely high energies estimated in the core of neutron stars are more than sufficient to create heavier particle species, beyond the traditional proton-neutron-electron admixture. 
It has become common in the literature to consider the entire spin-1/2 baryon octet \citep[e.g.][]{Glendenning:1982nc,Prakash:1995uw,Baldo:1999rq,Colucci:2014wda,Gomes:2014aka,Bhowmick:2014pma,Oertel:2014qza,Vidana:2015rsa,Chatterjee:2015pua,marquez17,Isaka:2017nuc,Roark:2018boj,Stone:2019blq,Sedrakian:2020kbi,Motta:2022nlj,Rather:2021azv,Huang:2022kej} but, recently, the role of the spin-3/2 decuplet has been slowly gaining attention, not just for its influence on the microscopic aspects of dense matter but also for the astrophysical implications, since its presence may reduce the radius and tidal deformability in intermediate mass neutron stars~\citep{Li:2018qaw,Yang:2018idi,Motta:2019ywl,Ribes:2019kno,Malfatti:2020onm,Sahoo:2020ipd,Li:2020ias,Raduta:2020fdn,backes2021effects,Thapa:2021syu,Dexheimer:2021sxs,Marczenko:2022hyt}. 
The lightest spin-3/2 baryons (the $\Delta$s) are only $\sim30\%$ heavier than the nucleons (protons and neutrons) and are even lighter than the heaviest spin-1/2 baryons of the octet (the $\Xi$s). Thus, unless the $\Delta$s are subject to a very repulsive coupling, they are expected to appear at the same density range as the hyperons (around 2 or 3 times the nuclear saturation density). Not much is known about how $\Delta$ baryons couple in dense matter, but their potential for isospin-symmetric matter at saturation density is expected to be attractive and in a range of $2$/$3$ to $1$ times the potential of the nucleons, which is of order $-80$ MeV  \citep{Drago:2014oja,Kolomeitsev:2016ptu,Raduta:2021xiz}. 

Additionally, it is of special interest to investigate how  spin-3/2 baryons are affected by the presence of strong  magnetic fields due to the possibility of them having large electric charge and additional spin and isospin projections. The effects of Landau levels in dense stellar matter containing $\Delta$ baryons was first discussed in the context of neutron-star matter by \citet{Thapa:2020ohp} and later by \citet{Dexheimer:2021sxs}. In this work we study for the first time the effects of strong magnetic fields in $\Delta$-admixed hypernuclear stellar matter, accounting for effects due to their anomalous magnetic moments (AMMs).

For magnetic fields larger than $\sim10^{16}$ G, the deformation of the stellar geometry away from spherical symmetry is above 2\%  \citep{Gomes:2019paw}. Therefore, the usual relativistic hydrostatic equations usually employed when describing non-magnetised stars, i.e., the Tolman-Oppenheimer-Volkoff equations \citep{Tolman:1939jz,Oppenheimer:1939ne}, which assume spherical symmetry as part of their derivation from the general relativity equations, cease to be adequate. 
For this reason, we make use of anisotropic solutions from the Einstein and Maxwell equations to explore for the first time the macroscopic structure of magnetars with strong internal magnetic fields and containing $\Delta$-admixed hypernuclear matter.
{Beyond accounting for the non-spherical configurations of stars and  anisotropies introduced by magnetic fields, this approach allows us to obtain an \textit{ab initio} magnetic field profile in the interior of a given star \citep{Dexheimer:2016yqu,Chatterjee:2018prm}.}

This work is structured as follows. In Section II, the formalism employed in the microscopic description of magnetized neutron-star matter is presented, as well as the procedure of going from the EoS to the macroscopic description of a compact star through General Relativity. In Section III, the results for the matter composition and stellar structure are shown and discussed, and, in Section IV, the main conclusions are drawn. 

\section{Formalism}\label{Formalism}

\subsection{Anomalous magnetic moment}

The AMM of a particle is a deviation from the magnetic moment of that particle, as predicted by the ``classical'' tree-level prediction. Historically, the term \textit{anomalous} was used to describe the deviation from the Dirac equation prediction for a system of fermions under the influence of a magnetic field, the magnetic moment, and thus, refers to fundamental particles. \textit{Dipole} moment, on the other hand, is used for composite particles, such as baryons, since their value depends on the configuration of quarks and gluons inside it, and thus, are not \textit{anomalous} in the strict sense. As commonly used in the literature and for simplicity, in this work we use the term AMM in all cases.

The energy spectrum of baryons with an AMM can be empirically determined, but a theoretical derivation of their values from first principles is yet an unaccomplished task.  The AMMs of nucleons are measured to a very high precision, with errors of the order of $10^{-9}$~\citep{PhysRevD.20.2139, doi:10.1126/science.aan0207}, but the same does not apply to heavier baryons. Measurements of the hyperon AMMs are precise to an order of $10^{-2}$ \citep{ParticleDataGroup:2020ssz}, 
while $\Delta$s are experimentally determined only for the positively charged $\Delta^{++}$ and $\Delta^+$. For the $\Delta^+$, there is a single measurement of $\mu_{\Delta^+}$/$\mu_N  = 2.7 ^{+ 1.0}_{-1.3} \pm 1.5$ that comes from the $\gamma p \rightarrow  p \pi \gamma'$  reaction~\citep{PhysRevLett.89.272001}, while for the $\Delta^{++}$ there are several measurements coming from the $\pi^+ p \rightarrow \pi ^+ p \gamma$ bremsstrahlung cross section, with values in the range $\mu_{\Delta^{++}}$/$\mu_N = 3.7 - 7.5$ ~\citep{ParticleDataGroup:2020ssz}. 
These measurements include systematic uncertainties, but additional theoretical uncertainties lead to errors $\sim\pm 3$. 
Complementary to experimental results, lattice quantum chromodynamics (LQCD) has been able to extract AMM values for $\Delta$ baryons.
The values utilized in this paper are based on the predictions from LQCD provided in \citep{CLOET2003157} that lie within the experimental uncertainties of the experimentally measured AMMs.

\begin{table}
    \centering
        \caption{Vacuum mass, electric charge, isospin $3^{\rm{rd}}$ component, spin, normalized magnetic moment, and normalized anomalous magnetic moment of baryons considered in this work.  Electric charge is shown in units of the electron charge and $\mu_N$ is the nuclear magneton.}
    \label{tab:baryons}
    \begin{tabular}{l c c c c c c }\hline
          & $M_b$ (MeV) & $q_b (e)$     & $I_{3\, b}$      &$S_b$      &$\mu_b/\mu_N$  & $\kappa_b/\mu_N$ \\ \hline
          $p$           & 939   & $+1$  & $+\frac{1}{2}$ &1/2    & $2.79$        & $1.79$\\              
          $n$           & 939   & 0     & $-\frac{1}{2}$ &1/2    & $-1.91$       & $-1.91$\\
          $\Lambda$     & 1116  & 0     & 0              &1/2    & $-0.61$       & $-0.61$\\ 
          $\Sigma^+$    & 1193  & $+1$  & $+1$           &1/2    & $2.46$        & $1.67$\\
          $\Sigma^0$    & 1193  & 0     & 0              &1/2    & $1.61$        & $1.61$\\
          $\Sigma^-$    & 1193  & $-1$  & $-1$           &1/2    & $-1.16$       & $-0.37$\\
          $\Xi^0$       & 1315  & 0     & $+\frac{1}{2}$ &1/2    & $-1.25$       & $-1.25$\\
          $\Xi^-$       & 1315  & $-1$  & $-\frac{1}{2}$ &1/2    & $-0.65$       & $0.06$\\
          $\Delta^{++}$ & 1232  & $+2$  & $+\frac{3}{2}$ &3/2    & $4.99$        & $3.47$\\
          $\Delta^{+}$  & 1232  & $+1$  & $+\frac{1}{2}$ &3/2    & $2.49$        & $1.73$\\
          $\Delta^{0}$  & 1232  & $0$   & $-\frac{1}{2}$ &3/2    & $0.06$        & $0.06$\\
          $\Delta^{-}$  & 1232  & $-1$  & $-\frac{3}{2}$ &3/2    & {$-2.45$}      &{$-1.69$}\\ \hline
    \end{tabular}
\end{table}

Different properties of baryons considered in this study are shown in Tab.~\ref{tab:baryons}. The AMM strength coefficients $\kappa_b$ are related to the magnetic moments $\mu_b$ through the relation
\begin{equation}
    \kappa_b=\mu_b - q_b\mu_N\frac{M_p}{M_b}~,
    \label{eq:kappa}
\end{equation}
which depends on the baryon charge $q_b$, the nuclear magneton  $\mu_N= e/2 M_p$, with $e$ being the electron charge, and the ratio of the proton mass $M_p$ to the baryon mass $M_b$. Although the expression \eqref{eq:kappa} is derived for spin-1/2 fermions in the non-relativistic regime, it is still commonly employed to the description of the spin-3/2 particles \citep[see][]{LOPEZCASTRO2002440,LOPEZCASTRO2001339}. This subject is controversial, as the Rarita-Schwinger equation with minimal coupling predicts a gyromagnetic ratio of 2/3, while low energy/optical theorems predict a value of 2. For a more in-depth discussion we refer to \citet{PhysRevD.62.105031}, which studies a generic non-minimal electromagnetic coupling in the Rarita-Schwinger formalism.
In this work, we also account for the leptons (electron and muon) AMMs, given by $\kappa_e/\mu_{B_e}= 1.15965 \times 10^{-3}$ and $\kappa_\mu/\mu_{B_\mu}= 1.16592 \times 10^{-3}$, respectively, with $\mu_{B_l}=e/2 M_l$, for $l=\{e,\mu\}$.

{\subsection{Matter description}}

To describe baryon-dense matter subject to a strong magnetic field, we must start from a Lagrangian density describing how the particles interact with each other and with the external electromagnetic field.
 The photons are simply described by the massless Proca Lagrangian density, followed by a term containing the electromagnetic interaction for charged baryons and leptons, and a term describing the AMMs of baryons $b$ and leptons $l$
\begin{align}
    \mathcal{L}_{EM}={}&-\frac{1}{4}F^{\mu \nu }F_{\mu \nu }\nonumber\\&+\sum_{b,l} \bar{\psi }_{b,l}\left ( -q_{b,l} A^\mu-\frac{1}{4}\kappa_b\sigma _{\mu \nu }F^{\mu \nu }\right )\psi _{b,l}~,
    \label{em}
\end{align}
where $F_{\mu \nu} = \partial_\mu A_\nu - \partial_\nu A_\mu$, $\sigma ^{\mu \nu }=\frac{i}{2}\left [ \gamma ^{\mu },\gamma ^{\nu } \right ]$, and the vector potential $A_\mu = (0, 0, Bx, 0)$ is chosen such that the magnetic field is parallel to the $z$-axis. Leptons are described additionally by the (free with respect to the strong force) Dirac Lagrangian density
\begin{equation}
    \mathcal{L}_{\rm lepton}=\sum_{l} \bar{\psi }_{l}\left [i \gamma _{\mu } \partial ^{\mu }-M_l \right ]\psi _{l}~.
\end{equation}

We make use of two different relativistic models to describe the still widely unknown strong interaction between baryons. 
The first model is a non-linear version of the Walecka model, where the baryon interactions are mediated by the $\sigma$, $\omega$, $\rho$ and $\phi$ mesons, within the mean field approximation. 
We choose the recently proposed L3$\omega\rho$ parametrization ~\citep{l3wr}, which includes an additional $\omega\rho$ interaction that allows the correct prediction of slope of the symmetry energy,  neutron-star radii and tidal deformabilities.
The $\phi$ meson {(with hidden strangeness)} couples only to the hyperons, allowing a higher maximum mass to be reproduced for neutron stars, thus circumventing the well-known hyperon puzzle \citep{Chamel:2013efa}, with an effect similar to the higher-order $\omega^4$ self interaction, also included. 
In addition to satisfying standard astrophysical constraints from LIGO/VIRGO and NICER, the model satisfies nuclear ground-state properties of finite nuclei and bulk properties of infinite nuclear matter. It is also consistent with the PREX-2 results for the symmetry energy of $L= 106 \pm 37$~\citep{PhysRevLett.126.172503}, though at the lower end of the error band.

The model Lagrangian density is written as $\mathcal{L} = \mathcal{L}_{b}+\mathcal{L}_{\rm m}$, where the first term is the (interacting) Dirac Lagrangian density for nucleons, hyperons, and $\Delta$s, and the second term accounts for the self interaction among scalar and vector mesons\footnote{Spin-3/2 baryons are in fact described by the Rarita-Schwinger Lagrangian density
\begin{equation}
    \mathcal{L}_{RS}=- \sum_{b=\Delta} \frac{1}{2} \bar{\psi }_{\mu\,  b}\left ( \epsilon ^{\mu \eta \lambda  \nu }\gamma _5\gamma _{\eta  }\partial _{\lambda  }-im_b\sigma ^{\mu \nu }\right )\psi _{\nu \, b}~,
\end{equation}
where $\epsilon ^{\mu \eta \lambda  \nu }$ {is the Levi-Civita symbol}, $\sigma ^{\mu \nu }=\frac{i}{2}\left [ \gamma ^{\mu },\gamma ^{\nu } \right ]$, and  $\psi _{\mu \, b}$ is a vector-valued spinor with additional components (when compared to the four component spinor in the Dirac equation). Nonetheless, its equation of motion can be written compactly as $\left( i \gamma^\nu \partial_\nu - m\right) \psi_\mu =0$, see \citet{dePaoli:2012eq}.}
\begin{align}
  \mathcal{L}_b = {}&\sum_{b} \bar{\psi }_{b} \left[ i \gamma _{\mu } \partial ^{\mu }
  - \gamma_0\left(g_{\omega b}\omega +g_{\rho b}{I }_{3b} \rho + g_{\phi b}\phi\right )-M_b^* \right ]\psi _{b}~,
 \end{align}
and
 \begin{align}
\mathcal{L}_{\rm m}={}&-\frac{1}{2} m_{\sigma }^{2}\sigma ^{2}
- \frac{\lambda }{3}\sigma  ^{3}-\frac{\kappa}{4}\sigma  ^{4}
+\frac{1}{2}m_{\omega }^{2}\omega^2+\frac{\xi}{4!}g_{\omega b}^4\omega^4  \nonumber\\
&
+\frac{1}{2}m_{\rho }^{2} \rho^2\,+g_{\omega \rho }\, \omega^2 \rho^2 
+\frac{1}{2}m_\phi^2\phi^2~,
\end{align}
where $I_{3 \, b}$ is the baryon isospin 3rd component, given in Table ~\ref{tab:baryons}. The mass of the baryons is modified by the medium, giving rise to an effective mass $M_b^*=M_b -g_{\sigma b}\sigma$. 
The fittings of the self couplings $\lambda$ and $\kappa$, and the couplings between the mesons $i=\{\sigma, \omega, \rho,\phi\}$ and baryons $b$, defined in terms of the ratios $x_{ib} = g_{i b}/g_{i N}$, are discussed in detail in \citet{l3wr}. 

Relevant for this work, the scalar meson couplings are fitted to reproduce the hyperon potential depth $U_{\Lambda}= -28$ MeV (for isospin-symmetric matter at saturation) and the remaining relative strength of the coupling constants are determined by SU(3) symmetry group arguments, as proposed by \citet{hyperonchi}, determining the complete hyperon-meson coupling scheme from a single free parameter, $\alpha_v$.
Despite the value of $\alpha_v$, hyperons are always present in the neutron-star matter and the sequence of hyperon thresholds are always the same, with an inversely proportional relationship between $\alpha_v$ and the stiffness of the EoS. 
In this work, we choose to use $\alpha_v=0.5$, which results in values for the additional potentials $U_{\Sigma}= +21.8$ MeV and $U_{\Xi}= +35.3$ MeV, and a stiffer EoS with respect to the values $\alpha_v = 0.75$ or 1.0 that are considered in \citet{hyperonchi}. Though the potential for the $\Xi^-$mesons is repulsive in the parametrization used, recent observational constraints predict it to be attractive~\citep{doi:10.1146/annurev-nucl-102419-034438, PhysRevLett.123.112002, Friedman:2021rhu}, but to reproduce such an attractive potential we would need an extra free parameter in the meson couplings~\cite{l3wr}.

The $\Delta$ couplings are treated more freely, as their behavior is not well known. The scarce information present in the literature, such as transport models~\citep{Cozma:2021tfu} and quasi-elastic scattering of electrons off nuclei~\citep{Bodek:2020wbk}, allows us to infer that the nucleon-$\Delta$ potential is slightly more attractive than the nucleon-nucleon one, so that, 
$U_{N}-30 {\rm ~MeV}\lesssim U_{\Delta}\lesssim U_N$,
which implies $x_{\sigma\Delta}$ is greater than $1$. Also, the vector coupling is constrained by LQCD results as respecting the relation $ 0 \lesssim
 x_{\sigma \Delta}- x_{\omega \Delta} \lesssim
 0.2$, and no constraint is put in the $x_{\rho\Delta}$ value \citep{WEHRBERGER1989797,Ribes:2019kno,Raduta:2021xiz}. Early investigations on the effect of these parameters were made in \citet{de2000delta,deltaslo2} and their role in the stellar particle composition and maximum-mass was studied considering $x_{\sigma \Delta} = 1.0$ and $1.1$, within two classes of relativistic mean-field models in \citet{Dexheimer:2021sxs}.
Following the previous study, we analyse the scenarios with {$x_{\sigma \Delta} = x_{\omega \Delta} = 1.0$ and $x_{\sigma \Delta} = x_{\omega \Delta} =1.2$}, keeping $x_{\rho \Delta} = 1.0$, that generates, respectively, potentials $U_\Delta=-66.25$ MeV (equal to the nucleon potential) and $-79.50$ MeV.

The second model we use in this work is the  chiral mean-field (CMF) model, which is based on a nonlinear realization of the chiral sigma model. As in all chiral models, the masses of the baryons are generated (and not just modified) by the medium. In this way, at large temperatures and/or densities they decrease allowing chiral symmetry to be restored, in agreement with LQCD results \citep{Aarts:2017rrl}. In this work, we
restrict ourselves to the hadronic version of the model (with leptons) developed by \citet{Dexheimer:2008ax}, and disregard the possibility of phase transitions to quark matter. We add an additional $\omega\rho$ interaction to be in better agreement with data for the slope of the symmetry energy, neutron-star radii, and tidal deformabilities \citep{Dexheimer:2015qha,Dexheimer:2018dhb}. We also add a higher-order $\omega^4$ interaction to reproduce more massive neutron stars \citep{Dexheimer:2020rlp}. 

The mean-field model Lagrangian density has the terms
\begin{equation}
\mathcal{L}_{b} = \sum_{b} \bar{\psi_b} [i\gamma_\mu\partial^\mu-\gamma_0 (g_{\omega b} \omega  +g_{\rho b}I_{3_b}\rho + g_{\phi b} \phi) - M_b^*] \psi_i~,
\end{equation}
and
\begin{align}
\mathcal{L}_{\rm m}={}& \frac{1}{2} \left(m_\omega^2 \omega^2 + m_\rho^2 \rho^2 + m_\phi^2 \phi^2\right)\nonumber \\
&+ g_4 \left(\omega^4 + \frac{\phi^4}{4} + 3 \omega^2 \phi^2 + \frac{4 \omega^3 \phi}{\sqrt{2}} + \frac{2 \omega \phi^3}{\sqrt{2}}\right)\nonumber \\
&- k_0 (\sigma^2 + \zeta^2 + \delta^2) - k_1 (\sigma^2 + \zeta^2 + \delta^2)^2\nonumber \\
&{}- k_2 \left(\frac{\sigma^4}{2}+\frac{\delta^4}{2} + 3 \sigma^2 \delta^2 + \zeta^4\right) - k_3 (\sigma^2 - \delta^2) \zeta \nonumber \\
&- k_4 \ \ln{\frac{(\sigma^2 - \delta^2) \zeta}{\sigma^2 \zeta}} -m_\pi^2 f_\pi \sigma&\nonumber \\
&{}- \left(\sqrt{2} m_k^ 2f_k - \frac{1}{\sqrt{2}} m_\pi^ 2 f_\pi\right) \zeta~,
\end{align}
where the effective mass of baryons is $M_b^*=g_{\sigma b} \sigma + g_{\delta b} I_{3\,b}\delta + g_{\zeta b} \zeta + M_{0_b}$, including additional corrections given by the scalar-isovector $\delta$ and scalar-isoscalar (with hidden strangeness) $\zeta$ mesons, and a small bare mass correction $M_0$. The couplings were fitted to reproduce hadronic vacuum masses, decay constants, nuclear saturation properties, and to reach more than $2.1$ M$_\odot$ stars. See  \citet{Roark:2018uls} for a complete list of coupling constants. We follow the SU(3) and SU(6) coupling schemes for the scalar and vector couplings of mesons and baryons. In this way, there are only two free parameters left: one fitted to reproduce for symmetric matter at saturation the potential $U_\Lambda=-27$ MeV and another one (${x_{\omega \Delta} }
=g_{\omega\Delta}/g_{\omega N}=1.25$) fitted to reproduce under the same conditions $U_\Delta=-64$ MeV $\sim U_N$. They result additionally in $U_\Sigma=6$
MeV and $U_\Xi=-17$ MeV. A much larger { $x_{\omega \Delta}$}
would suppress all $\Delta$s, while a much lower value would suppress all hyperons.

For both models, the equations of motion for the mesonic fields are obtained from the Lagrangian densities via the Euler-Lagrange equations. 
Under charge neutrality and $\beta$-equilibrium conditions, we can write the chemical potential of a baryon as a relation between the chemical potential of the neutron and the electron, $\mu_n$ and $\mu_e$, respectively, and its electric charge, i.e.,
\begin{equation}
\mu_b = \mu_n - q_b \mu_e~.
\end{equation}
At low (effectively zero) temperature, the Fermi energy spectra of baryons is 
\begin{equation}
    E_{F \, b}^\ast=\mu_b-g_{\omega b}\omega- g_{\rho b} I_{3 b} \rho- g_{\phi b} \phi~,
\end{equation}
while for leptons it is simply $E_{F \, l}^\ast=\mu_e$. 

In the presence of a magnetic field, the Fermi momentum (squared) can be calculated from the difference between the Fermi energy (squared) and
\begin{enumerate}
    \item the square of the effective mass modified by the AMM for particles that are not electrically charged $(q_b = 0)$,
\begin{equation}    
    k_{F, b}^2 (s)  = {E_{F\, b}^\ast}^2 - \left( M_b^{\ast} - s \kappa_b B \right)^2 ;
     \label{10}
\end{equation}
    \item the square of the effective mass modified by Landau quantization and AMM for particles that are electrically charged $(q_b \neq 0)$,
\begin{equation}
    k_{F , b}^2 (\nu, s)  =  {E_{F\, b}^\ast}^2 - \left( \sqrt{ M_b^{\ast 2}+2 \nu |q_b| B } - s \kappa_b B \right)^2.
    \label{11}
\end{equation}
\end{enumerate}
For the momentum of leptons, $M^*$ is simply $M$. In the latter case, the Fermi momentum refers to the local direction of the magnetic field, hereafter defined as the $z$-axis. In the transverse direction to the local magnetic field, the Fermi momentum is restricted to discrete values $2\nu|q_b|B$, where the Landau levels $\nu$ relate to the orbital angular momentum $n$ via the relation
\begin{equation}
    \nu=n +\frac{1}{2}-\frac{s}{2}\frac{q_b}{|q_b|}~,
    \label{nu}
\end{equation}
where  $n=0, 1, 2...$ . For particles with spin $1/2$, the first Landau level ($\nu=0$) is occupied by a single spin projection: $s=+1$ for $q_b>0$ and $s=-1$ for $q_b<0$. The second level ($\nu=1$) and above {are} occupied by both spin projections $s=\{\pm 1\}$. For the spin-3/2 positively charged $\Delta$s, the first level ($\nu=0$) is occupied by the spin projections s$=\{+3, +1\}$, the second level ($\nu=1$) by $s=\{+3, \pm 1\}$, and hereafter all spin states are occupied. For the negatively charged $\Delta^-$ spin projection, signs are reversed for the lowest levels.

At zero temperature, there is a maximum Landau level allowed, beyond which the momentum in Eq.~\eqref{11} becomes imaginary given by 
\begin{equation}
    {\nu_{\rm max}}_b (s) = \left\lfloor\frac{\left( E_{F\, b}^\ast + s \kappa_b B\right)^2 - {M_b^\ast}^2} {2 |q_b| B}\right\rfloor~.
\end{equation}
The number density for each baryon is also defined separately for electrically charged and uncharged particles, respectively, 
\begin{align}
 n_{b}^{(q_b\neq 0)}&={}\frac{|q_b|B}{2\pi^2}\sum_{\nu, s} {k_F}_b(\nu,s)~;\label{eq:dens_nocharge}
 \\
 n_b^{(q_b = 0)}&={}\bar{\psi}_b\psi_b=\frac{1}{2\pi^2}\sum_s \left\{\frac{k_{F\,b}^3(s)}{3} - \frac{s\kappa_b B}{2} \right. \nonumber\\
 &{} \Bigg{[} \left(M^\ast_b - s \kappa_b B\right){k_F}_b(s) + \nonumber \\
 &\left.E_{F\, b}^{\ast \, 2}\left(\arcsin{\left(\frac{ M^\ast_b - s \kappa_b B}{ E_{F \,b}^\ast}\right)}-\frac{\pi}{2}\right)\bigg{]}\right\}~,
\end{align}
as well as the scalar densities, 
\begin{align}
    n_{s\, b}^{(q_b \neq 0)}={}&\bar{\psi}_b\gamma_0\psi_b=\frac{|q_b|B M_b^\ast}{2\pi^2}\sum_{s, \nu} \frac{\sqrt{{M_b^\ast}^2+2\nu|q_b|B} - s \kappa_b B}{\sqrt{{M_b^\ast}^2+2\nu|q_b|B}}\nonumber\\
    &\times\ln\left|\frac{{k_F}_b(\nu,s)+E_{F\,b}^\ast}{\sqrt{{M_b^\ast}^2+2\nu|q_b|B} - s \kappa_b B}\right|~;\\
n_{s\, b}^{(q_b = 0)}={}&\frac{M_b^\ast}{4\pi^2}\sum_s\left[\vphantom{\frac{ M^\ast_b }{ E_{F \,b}^\ast}} 
    E_{F\,b}^\ast {k_F}_b(s)\right.\nonumber\\
    &{}-\left. {\left( M^\ast_b - s \kappa_b B\right)^2}\ln\left|\frac{k_{F\,b}(s)+E_{F\,b}^\ast}{ M^\ast_b - s \kappa_b B}\right|\right] .
\end{align}

The expressions for energy density and pressure are different for each model and can be obtained from the energy-momentum tensor for matter (discussed in the following).

\subsection{Macroscopic structure}
\label{sec:lorene}

For spherically symmetric neutron stars, given an EoS $P(\varepsilon)$, the global structure can be obtained by solving the Tolman-Oppenheimer-Volkoff (TOV) equations of hydrostatic equilibrium
\begin{align}
\frac{d M}{d r} ={}& 4 \pi r^2 \varepsilon(r)~, \label{eq:TOVi} \\
\frac{d \nu}{d r} ={}& \left( M(r) + 4 \pi r^3 P(r) \right) \frac{1}{r^2}  \left( 1-\frac{2 M}{r} \right)^{-1}~,  \\ 
\frac{d P}{d r} ={}& -\left( \varepsilon(r) + P(r) \right) \frac{d \nu}{d r}~, \label{eq:TOVf}
\end{align}
where $M(r)$ is the stellar mass contained within the radius $r$ and $\nu(r)$ is a gravitational potential for the line element in spherical coordinates
\begin{equation}
ds^2 = - e^{2 \nu(r)} dt^2 + \left( 1 - \frac{2 M}{r} \right)^{-1} dr^2 + r^2 (d \theta^2 + \sin^2 \theta d \phi^2)~.\label{eq:linetov}
\end{equation}

\begin{figure*}
    \centering
    \includegraphics[width=\linewidth]{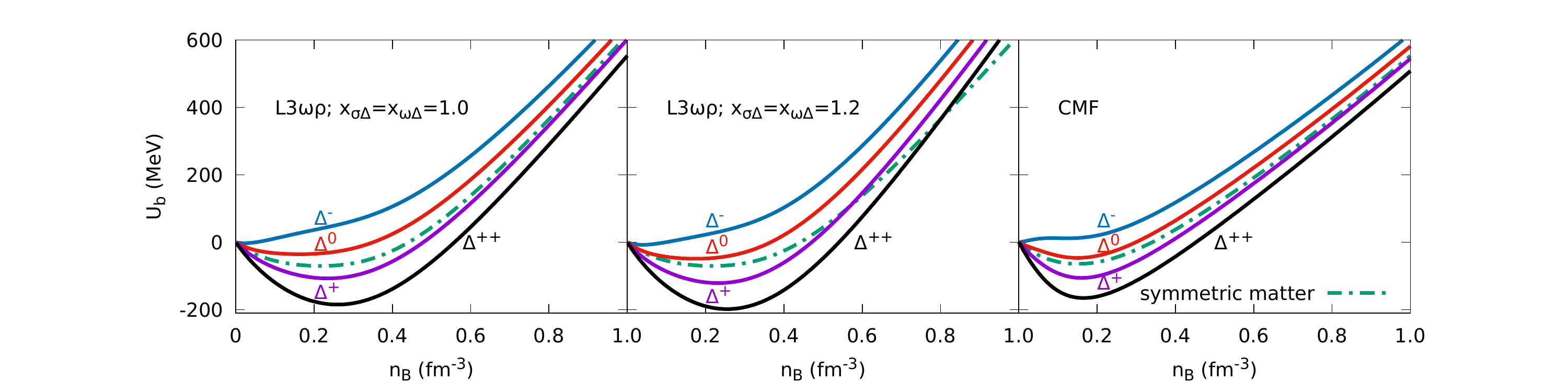}
    \vspace{-0.3cm}
    \caption{Single-particle potentials of $\Delta$ baryons as a function of baryon number density in isospin-symmetric nuclear matter (dashed-dotted line) and pure neutron matter (solid lines) for the L3$\omega\rho$ model using two different $\Delta$ scalar interaction strengths (left and middle panels) and for the CMF model (right panel).}
    \label{fig:potential}
\end{figure*}

The TOV equations cannot be applied to describe the structure of the magnetars we study in this work because the spherical symmetries assumed in Eq.~\eqref{eq:linetov} will not hold. This is due to the strong magnetic fields we infer for such objects, which produce highly deformed stellar shapes. 
Instead, the stellar structure must be determined by solving equations in General Relativity describing the stationary configuration for the fluid, coupled with Einstein field equations. The energy-momentum tensor, which contains the information on the matter properties of stars, enters the stellar structure equations as the source of the Einstein equations. 
Neglecting the coupling to the electromagnetic field, one generally assumes a perfect fluid and the energy-momentum tensor takes the form
\begin{equation}
T_f^{\mu\nu} = (\varepsilon + P)\; u^\mu u^\nu + P\; g^{\mu\nu}~,
\label{eq:perfectfluid}
\end{equation}
where $\varepsilon$ denotes the (matter) energy density, $P$ the
pressure, and $u^\mu$ the fluid four-velocity. 

The EoS then relates pressure and energy density to the relevant thermodynamic quantities. In \citet{Chatterjee2015}, the general expression for the energy-momentum tensor in the presence of an electromagnetic field was derived, starting from a microscopic Lagrangian including interactions between matter and the electromagnetic field
\begin{align}
T^{\mu\nu}={}&T_f^{\mu\nu} 
 + \frac{1}{\mu_0} \left( - B^\mu B^\nu + (B\cdot B) u^\mu u^\nu + \frac{1}{2}
   g^{\mu\nu} (B \cdot B) \right) \nonumber \\
 &{}+ \frac{x}{\mu_0}  \left( B^\mu B^\nu - (B\cdot B)( u^\mu u^\nu + g^{\mu\nu})\right) ~,
\end{align}
where $\mu_0$ is the vacuum permeability, $g^{\mu\nu}$ the metric tensor, and $x$ is the magnetisation.
The electromagnetic field tensor has been expressed as
$F_{\mu\nu} = \epsilon_{\alpha\beta\mu\nu} u^{\beta} B^{\alpha}~,$
with $\epsilon_{\alpha\beta\mu\nu}$ being the four-dimensional Levi-Civita symbol \citep{Gourgoulhon}. Assuming an isotropic medium and a magnetisation parallel to the magnetic field, the magnetisation tensor $M_{\mu \nu}$ can be written as 
\begin{equation}
M_{\mu \nu} =  \epsilon_{\alpha \beta \mu \nu} u^{\beta} a^{\alpha}~,
\label{eq:magtensor}
\end{equation}
with the magnetization four-vector defined as
$a_{\mu} = \frac{x}{\mu_0} B_{\mu}$.
In the absence of magnetisation, i.e. for $x=0$, this expression
reduces to the standard magnetohydrodynamics form for the energy-momentum tensor \citep[c.f.][]{Gourgoulhon}.

Strong magnetic fields result in an anisotropy of the energy momentum tensor and break spherical symmetry, such that with increasing strength of the magnetic field, the shape of a magnetar departs more and more from a spherical shape. Interpreting the spatial elements of the fluid rest frame energy-momentum tensor as pressures, then there is a difference induced by the orientation of the magnetic field, commonly referred to as ``parallel'' and ``perpendicular'' pressures.
Several earlier works tried to compute the mass-radius relations of strongly magnetised neutron stars through a first approach using isotropic TOV equations
\citep{Rabhi,Ferrer,Strickland,Lopes,Dexheimer2014,Casali}. In these works, the components of the macroscopic energy-momentum tensor in the fluid rest frame are used to obtain the energy density $\varepsilon$, parallel ($P_{\parallel}$) and perpendicular ($P_{\perp}$) pressures. In Heaviside-Lorentz natural units, the pure electromagnetic contribution to the energy-momentum tensor, which is anisotropic, has values of $B^2/2$ and $-B^2/2$ in the perpendicular and parallel  directions to the local magnetic field, respectively. However, this approach can drastically overestimate the mass of neutron stars \citep[as shown in Fig.~3 of][]{Gomes:2017zkc}. 

Several works obtained the global structure models of magnetars by solving coupled Einstein-Maxwell equations, taking into account the anisotropy of the stress-energy tensor  \citep{1995A&A...301..757B,Cardall:2000bs,Oron,Ioka,Kiuchi2008,Yasutake,Frieben:2012dz,Yoshida,Bucciantini,Thapa:2020ohp,Tsokaros:2021pkh}. In these studies either a perfect fluid, a polytropic EoS, or a realistic EoS was assumed, but do not take into account the magnetic field modifications due to its quantisation.

Ideally, to explore magnetic field effects such as Landau quantisation and AMM on the global properties of the star, one must solve the coupled Einstein-Maxwell equations, along with a magnetic field dependent EoS. In ~\citet{Chatterjee2015} and \citet{Franzon:2015sya}, global numerical models for magnetars were obtained by consistently solving Einstein-Maxwell equations with magnetic field dependent quark EoS. It was however explicitly demonstrated by \citet{Chatterjee2015,Chatterjee2021} that the maximum mass of a neutron star is minimally modified due to the magnetic field dependence of the microscopic EoS, even for the highest magnetic fields. Therefore in this work, we assume a non-magnetic ($B=0$) matter contribution to the EoS to compute global neutron-star models and the magnetic field enters structure calculations only through the dominant pure electromagnetic field contribution. 
Although it remains to be checked explicitly in future work, the effects of Landau quantisation and AMMs are not expected to sensibly affect the results of this study. Note however, that this is not the case for microscopic properties of matter, as discussed in the following.

\section{Results}

\subsection{Matter properties}
\label{sec:eos}

We start our discussion with the single-particle interaction potential for the $\Delta$ baryons in dense nuclear matter, which is a measure of the energy cost of adding one particle $b$ in a $b$-less matter with the given condition. For the L3$\omega\rho$ model, it can be written as
\begin{equation}
U_b=-g_{\sigma b}\sigma +g_{\omega b}\omega+g_{\rho b} I_{3b} \rho +g_{\phi b} \phi~,
\end{equation}
and, for the CMF model,
\begin{equation}
U_b=g_{\sigma b} \sigma + g_{\delta b} I_{3_b}\delta + g_{\zeta b} \zeta-m_{b,vac}+g_{\omega b}\omega+ g_{\rho b} I_{3b} \rho +g_{\phi b} \phi~.
\end{equation}
 In isospin-symmetric nuclear matter, all families of baryons experience the same potential, since the meson field $\rho$ (and $\delta$) are null in this situation. In neutron rich matter, particles with positive isospin projections (as the positively charged $\Delta$s) are more bound than their zero- and negative-isospin counterparts, with the largest difference occurring for pure neutron matter, that can be taken as an extrapolation of neutron-star matter in $\beta$-equilibrium.
 The first two panels of Fig.~\ref{fig:potential} show how the L3$\omega\rho$ model scalar and vector interactions affect the $\Delta$ potentials. In all cases, the particle potentials eventually become positive as the density increases, corresponding to unbound states, but they stay negative in the relevant interval of densities around nuclear saturation, where their depth determines how much they are bound. 
 
 For the L3$\omega\rho$ model, the  larger the scalar coupling value (i.e., the parameter $x_{\sigma \Delta}$), the lower the potentials are in the low density regions.
 Complementary to that, the larger the vector coupling (i.e., the parameter $x_{\omega\Delta}$), the more repulsive the potentials for $\Delta$s are, which reflect in more positive curves in the high density region, where the repulsive channel dominates.
 Also, it can be seen that the potential depends less on the species of $\Delta$ in the CMF model. 
 {Magnetic-field effects are not included, but it was verified that fields up to $B=3\times10^{18}$ G do not affect these results.}

For $B=0$, the matter EoS (namely, $P$ vs. $\varepsilon$) shows a simple monotonically increasing behavior, however its derivatives show interesting features generated, e.g., by changes in particle composition. 
Next, we discuss the incompressibility modulus (usually referred to simply as compressibility), given by
\begin{equation}
    K=9\frac{\partial P}{\partial n_B}~.
\end{equation}
At saturation density, compressibility values for isospin-symmetric matter can be compared with laboratory data. We find values of 256 MeV and $300$ MeV for the L3$\omega\rho$  and CMF models, respectively. Laboratory values range between $220<K<260$ MeV \citep{dutra2014, l3wr} and $250<K<315$ MeV \citep{Stone:2014wza}. 

\begin{figure}
    \centering
    \includegraphics[width=\linewidth]{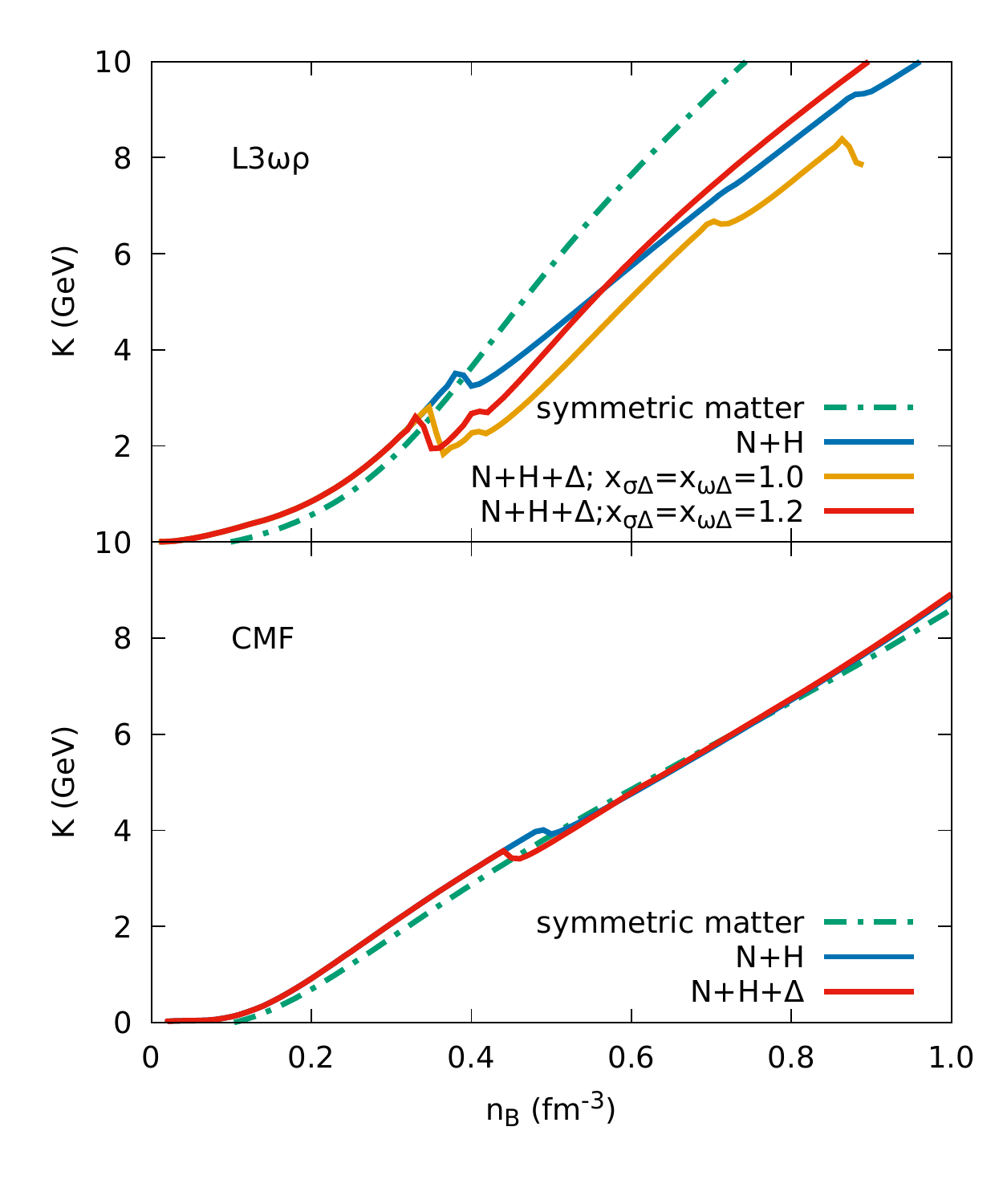}  
    \vspace{-0.8cm}
    \caption{Compressibility as a function of baryon number density in isospin-symmetric matter with only nucleons (dashed-dotted line) and neutron-star matter (full lines) shown for different compositions and interaction strengths for the L3$\omega\rho$ (top panel) and for the CMF model (bottom panel). $B=0$.}
    \label{fig:compress}
\end{figure}

\begin{figure}
    \centering
    \includegraphics[width=\linewidth]{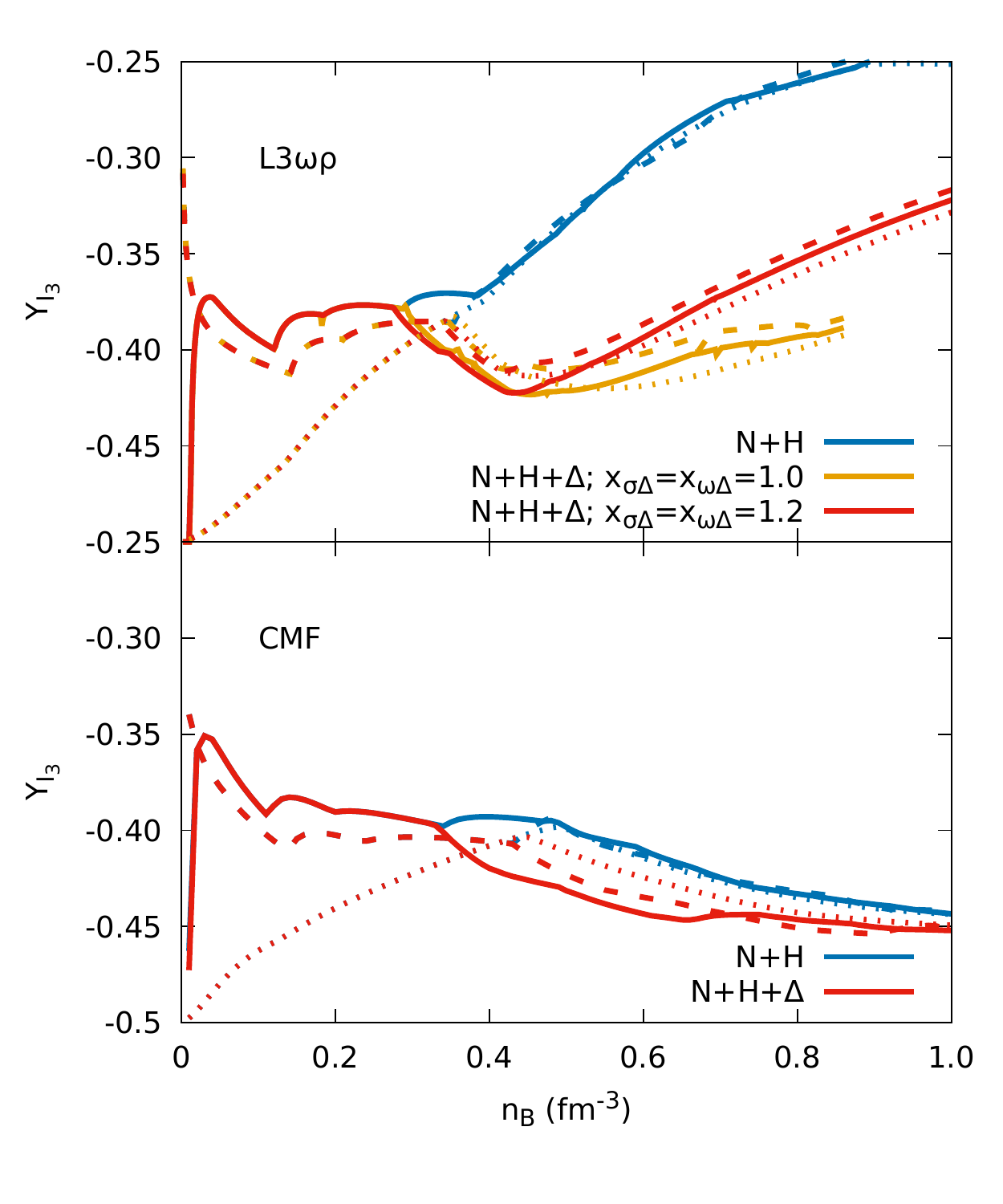}
    \vspace{-0.8cm}
    \caption{Isospin fraction as a function of baryon number density for neutron-star matter with $B=0$ (dotted lines) and with magnetic field $B=3\times 10^{18}$ G when considering (solid lines) or disregarding (dashed lines) the effects of the anomalous magnetic moments and shown for different compositions and interaction strengths. The top and bottom panels show results for the L3$\omega\rho$ and CMF models, respectively.}
    \label{fig:rhoiz}
\end{figure}

The top panel of Fig.~\ref{fig:compress} shows the effect of the inclusion of different particle species in the compressibility for the L3$\omega\rho$ model, in the absence of an external magnetic field. The kinks in the curves are consequence of the onset of new particle species, which are shifted to lower densities by the inclusion of both $\Delta$s and respective stronger scalar interactions. {For $x_{\sigma \Delta} = 1$, the effective mass of nucleons becomes zero at $n_B \sim 0.85$ fm$^{-3}$ and, for this reason we lack solutions at higher densities.}
The bottom panel of Fig.~\ref{fig:compress} shows that in the CMF model the kinks are much smaller than in the L3$\omega\rho$ model, with the only displacement of the curve occurring at the onset of the first non-nucleon baryon. As a consequence, the different CMF EoSs behave more similarly as the density increases.

\begin{figure*}
    \centering
    \includegraphics[width=\linewidth]{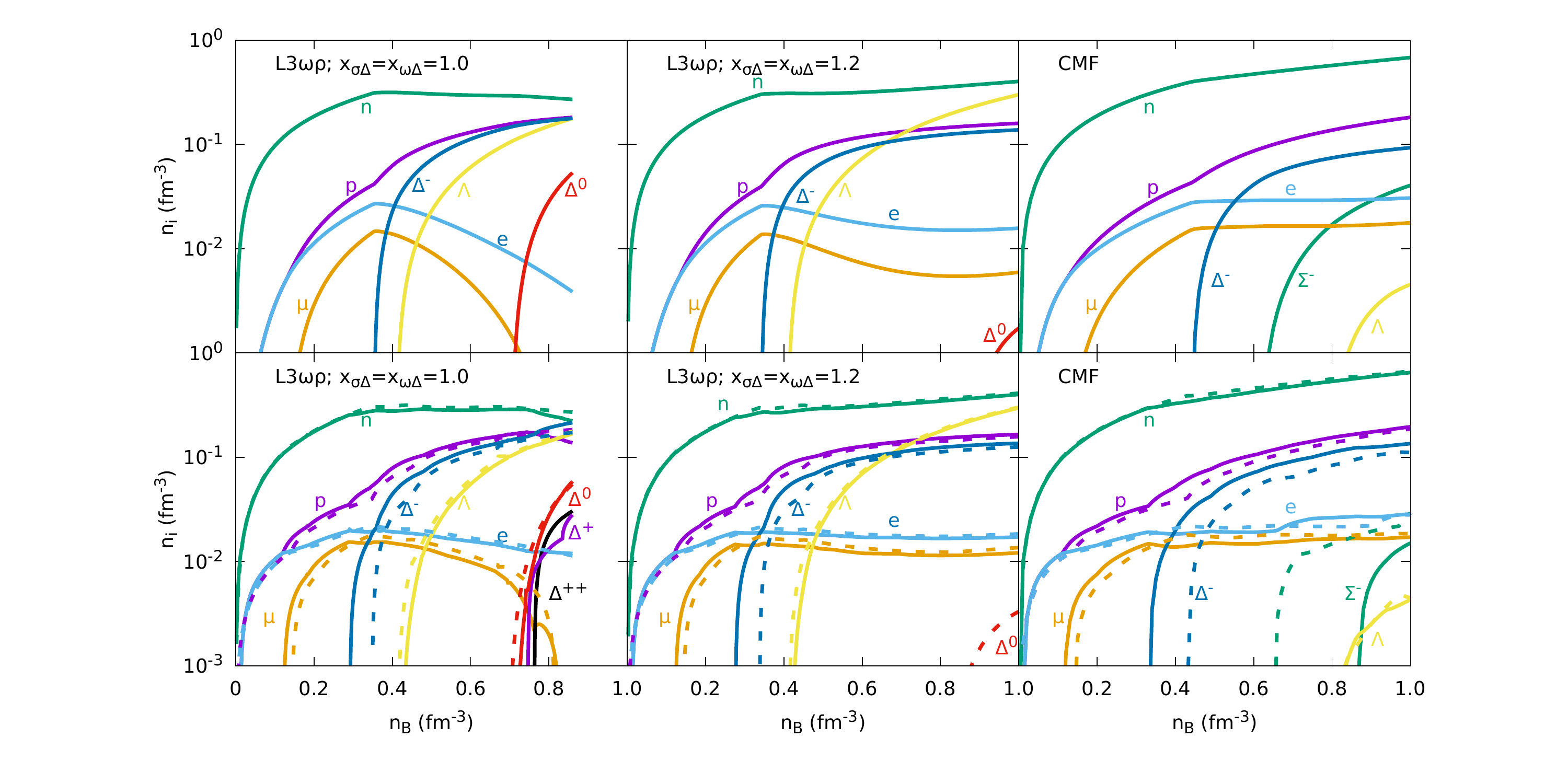}
    \vspace{-0.8cm}
    \caption{Particle composition of neutron-star matter with $\Delta$s, with $B=0$ (top panels) and magnetic field $B=3\times 10^{18}$ G (bottom panels), when considering (solid lines) or disregarding (dashed lines) the effects of the anomalous magnetic moment. The left and middle panels show results for the L3$\omega\rho$ model with different interactions, while the right panel shows results for the CMF model.}
    \label{fig:dens_nyd}
\end{figure*}

\begin{figure}
    \centering
    \includegraphics[width=\linewidth]{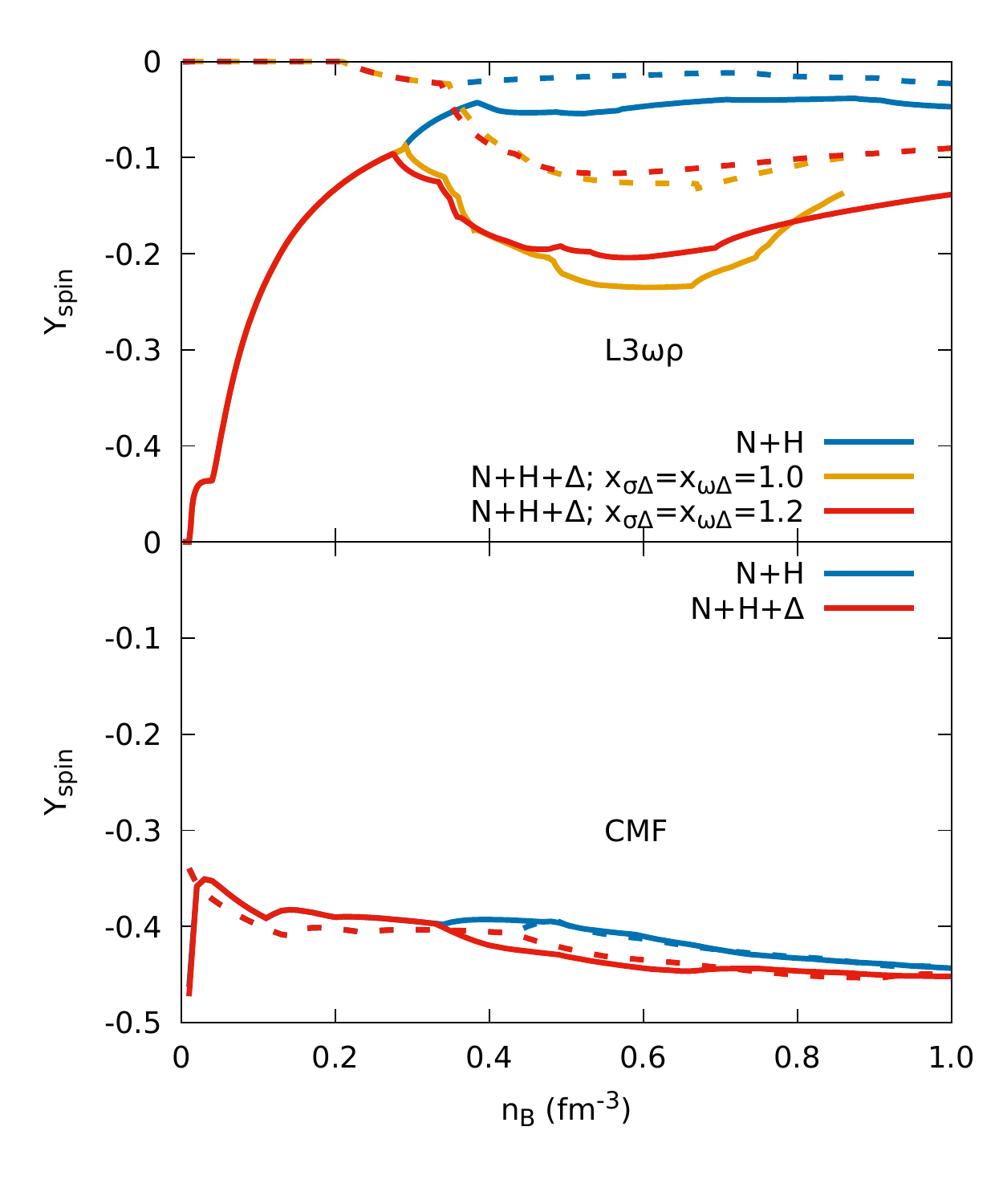}
    \vspace{-0.8cm}
    \caption{Spin polarization fraction as a function of baryon number density for neutron-star matter with magnetic field $B=3\times 10^{18}$ G, when considering (solid lines) or disregarding (dashed lines) the effects of the anomalous magnetic moments and shown for different compositions and interaction strengths. The top and bottom panels show results for the L3$\omega\rho$ and CMF models, respectively.}
    \label{fig:rhospin}
\end{figure}

The stiffer EoSs are formally the ones with larger values of the speed of sound $v_s$, but here we discuss stiffness with respect to $K$, related to $v_s$ through $v_s^2=K/(9\mu)$ (\citep{Dexheimer:2007mt}). Isospin symmetric matter is softer at low densities, but becomes stiffer at large densities due to the Pauli exclusion principle because, as only nucleonic matter is considered, higher Fermi levels must be occupied (see Fig.~2). 
The behavior of neutron-star matter (charge neutral and in chemical equilibrium) depends on the composition, but it is always softer than the symmetric matter case after the hyperon or $\Delta$ onsets, as the presence of new Fermi levels turns the EoS softer.
Matter with hyperons but no $\Delta$s is stiffer at intermediate densities (than matter with $\Delta$s), however it is softer at large densities, especially in the case of strong scalar interaction (for the L3$\omega\rho$ model). This trend was noticed previously by \citet{Dexheimer:2021sxs}, where we showed that the inclusion of $\Delta$s could turn the EoS stiffer (than the cases where they were absent), despite the fact that the new degrees of freedom soften the EoS. This is related to isospin asymmetry, which we discuss in the following.

\begin{figure*}h!
    \centering
    \includegraphics[width=\linewidth]{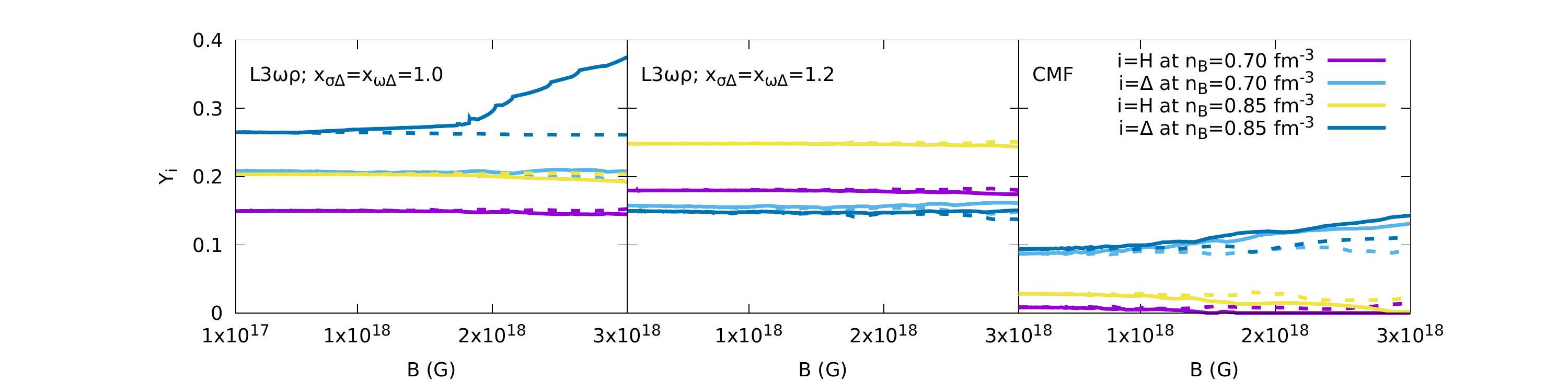}
    \vspace{-0.3cm}
    \caption{Exotic particle (hyperon and $\Delta$) fraction as a function of magnetic field strength when considering (solid lines) or disregarding (dashed lines) the effects of the anomalous magnetic moment, shown for two given densities. The left and right panels show results for the L3$\omega\rho$ with different interactions and CMF models, respectively.}
    \label{fig:frac_B_nyd}
\end{figure*}

We define the isospin fraction as the average 3$^{\rm rd}$ isospin component of a given matter composition, weighted by the relative densities, i.e.,
\begin{equation}
    Y _{I_3}=\frac{\sum_b I_{3\,b}n_b}{\sum_b n_b}~,\label{eq:rhoiz}
\end{equation}
as shown in Fig.~\ref{fig:rhoiz}. For nucleonic matter only, $Y _{I_3}=0$ means matter with the same amount of protons and neutrons, while $Y _{I_3}=-0.5$ means pure neutron matter. The density at which the curves with and without $\Delta$s split marks the appearance of the $\Delta^-$s, which increase the isospin asymmetry (turn the isospin fraction more negative). 
The effect is much larger for the L3$\omega\rho$  model (top panel) than the CMF model (bottom panel), which hints that the amount of $\Delta$s reproduced in each model is different. Both effects generated by the magnetic fields, i.e. Landau quantization and AMM, decrease the isospin asymmetry (less negative $Y _{I_3}$) at low and intermediate densities.

A better understanding of the effects of the inclusion of $\Delta$ baryons, magnetic fields, and AMM in neutron-star matter subject to strong magnetic fields can be obtained from Fig.~\ref{fig:dens_nyd}.
Comparing the top row ($B=0$) with the lower one ($B=3\times 10^{18}$ G), we can see that some of the charged particles are favored when magnetic field effects without AMMs are considered, an effect that is more pronounced for protons, whose onset density is pulled to very low densities for both models. 
As a consequence, their population becomes more similar to the neutron one in densities below $\sim 0.05$ fm$^{-3}$, turning $Y _{I_3}$ less negative. The inclusion of AMM enhances this effect. This explains why the isospin asymmetry depends both on the magnetic field and on the AMM in the lower density region, as shown in Fig.~\ref{fig:rhoiz}.
The $\Delta^-$ threshold (at densities around $0.3$ fm$^{-3}$) coincides with the region at intermediate densities beyond which the N+H+$\Delta$ EoSs are softer than the respective N+H EoSs. The $\Lambda$ (and the $\Sigma$ in the CMF model) hyperons appear at larger densities than the $\Delta^-$s. The remaining $\Delta$s appear at much larger densities and in amounts that depend on the interactions in the L3$\omega\rho$  model.

To discriminate AMM effects on the particle composition is not trivial, as they depend on the AMM coupling strength and sign, on the particle mass, charge, and density. Additionally,  different spin projections are separately enhanced or suppressed, but this cannot be clearly seen in Fig.~\ref{fig:dens_nyd}, as it follows the usual convention and shows the sum of all spin projections for each particle. For this reason, we make use of a quantity that reveals the degree of spin polarization, more suited to discuss spin projection asymmetry of fermions.

We define the total spin polarization of a given matter composition, weighted by the relative densities, in analogy to Eq.~\eqref{eq:rhoiz}, i.e.,
\begin{equation}
    Y_{\rm{spin}}=\frac{\sum_{b,s} s n_{b}(s)}{\sum_{b,s} n_{b}(s)}~,
\end{equation}
and shown the results in Fig.~\ref{fig:rhospin}.
For a fixed magnetic field strength, all charged particles are fully spin polarized at low densities: only spin projection 1/2 for protons and spin projection 3/2 for positive $\Delta$s, only spin projection -1/2 for leptons and negative $\Sigma$s, and spin projection -3/2 for negative $\Delta$s.
When AMMs are considered, neutral particles obey the same logic, presenting only positive (negative) spin projections according to their positive (negative) sign of $\kappa_b$.
At intermediate densities, full polarization is broken for more massive particles, but not for leptons and $\Lambda$s. 
But, regardless, the polarization never goes to zero, meaning that partial spin projection imbalance remains at high densities. Overall, spin polarization fraction is much stronger for the CMF model (bottom panel) than for the L3$\omega\rho$  model (top panel).
Full polarization can be understood from Eqs.~\eqref{10}, \eqref{11}, and \eqref{nu}, which explains why particles with different isospin projections present different momenta and why particles occupying the first Landau level ($\nu=0$) are more abundant when only a few levels are occupied. This happens for strong magnetic fields and low particle densities, or simply less massive particles.

It is a well-established concept that the magnetic field is not constant within neutron stars, but increases towards their centers where the density is larger. But, before we discuss stellar configurations with macroscopic magnetic fields in detail, we study how one more relevant quantity changes as a function of magnetic field strength. The fraction of exotic particles can be defined as the following quantity
\begin{equation}
Y_{i}= \frac{\sum_{b\in i}n_{b}}{\sum_b n_b}~,
\end{equation}
for $i=H$ or $\Delta$, shown in Fig.~\ref{fig:frac_B_nyd}. On the left panel for the L3$\omega\rho$  model, the amount of $\Delta$s is slightly reduced at a given density but then increases tremendously at the larger density when the AMM is included, a  behaviour quantitatively not reproduced with larger coupling constants, as seen on the middle panel.
The amount of hyperons, on the other hand, is not significantly modified by the magnetic field, only slightly decreases in the presence of AMM and is affected by the small fluctuations related to the De Haas-Van Alphen oscillations \citep{VanHalfen}. 
The right panel shows the same qualitative behavior for the CMF model, which has a more clear substitution of hyperons in favor of deltas for higher values of $B$, independently of the density or accounting for the AMM.

\subsection{Macroscopic structure}

 \begin{figure}
    \centering
    \includegraphics[width=\linewidth]{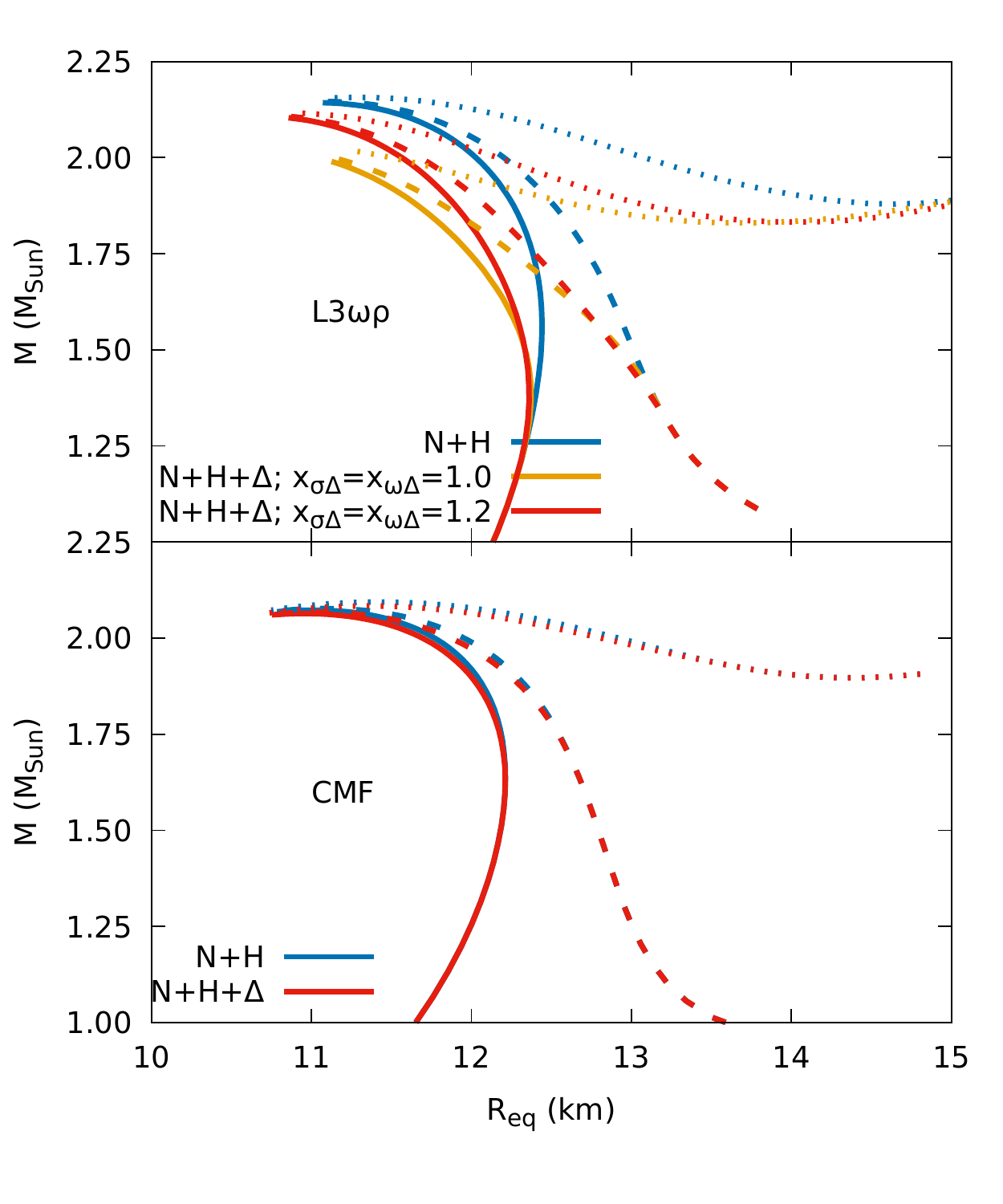}
    \vspace{-0.8cm}
    \caption{Stellar mass as a function of equatorial radius for different compositions and interaction strengths, for central magnetic fields $B = 0$ (solid lines), $B = 5\times 10^{17}$ G (dashed lines), and $B = 10^{18}$ G (dotted lines). The top and bottom panels show results for the L3$\omega\rho$ and CMF models, respectively.}
    \label{fig:mr}
\end{figure}
 
In a previous study, \citet{Dexheimer:2021sxs} obtained results on the effects of the inclusion of $\Delta$s in neutron stars for both the $L3\omega\rho$ and CMF models without magnetic fields, using standard TOV equations. 
As expected from the discussion regarding the hyperon puzzle \citep{Chamel:2013efa}, pure nucleonic stars are always the most massive configuration, and the inclusion of hyperons decreases the maximum mass obtained. 
It was noticed that the inclusion of $\Delta$s, however, does not modify significantly (and in some cases increases) the maximum stellar  mass in relation to the composition with only nucleons and hyperons, and this effect is more obvious for the cases where $\Delta$s are more abundant. It is argued that the addition of $\Delta$s decreases the fraction of nucleons (in fact, neutrons) and hyperons (mostly $\Lambda$s) to create $\Delta$s and some protons, in a way that the overall increase in isospin asymmetry turns the EoS stiffer, even when more species are present.

As described here in section~\ref{sec:lorene}, to compute the effect of the strong magnetic fields on the structure of the magnetars, one must solve the coupled Einstein–Maxwell equations with the equations of state (described in section~\ref{sec:eos}). For the chosen poloidal field geometry, we solve the Einstein–Maxwell equations within the numerical library LORENE\footnote{\url{http://www.lorene.obspm.fr}} using a multi-domain spectral method. In Fig~\ref{fig:mr}, we show the mass radius relations for the $L3\omega\rho$ and the CMF models, with and without $\Delta$s, as a function of equatorial radius for sequences of constant stellar central magnetic field.
Despite the fact that, for the choices discussed here for $x_{\sigma \Delta}$ and $x_{\omega\Delta}$ parameters in the $L3\omega\rho$ model, the masses of N+H+$\Delta$ stars never surpass the respective N+H configurations, we still observe that the the maximum mass (shown in Fig.~\ref{fig:mr}) follows the same ordering of a large (and most relevant) portion of Fig.~\ref{fig:rhoiz} for the compressibility.

In Fig.~\ref{fig:mr}, any differences between the mass-radius curves for the  $B=0$ case (solid lines) arise from the differences in the (non-magnetic) EoS, while the differences with magnetic field come from the pure electromagnetic field contribution.
We know that the Lorentz force originating from the pure electromagnetic field affects the low density part of the EoS. This is why the maximum mass of very massive stars does not change with increasing magnetic field strength, but the mass and radius of less massive stars increase significantly. For the $L3\omega\rho$ model, the inclusion of $\Delta$s decreases modestly the maximum stellar mass, especially for the larger coupling. However, for the CMF model, we do not see meaningful changes on the mass-radius diagram with the inclusion of $\Delta$s. 
From Table~\ref{tab:centralvalue} we see that, keeping the radius of the neutron star fixed (going up vertically in Fig.~\ref{fig:mr}), the increase in the strength of the central magnetic field increases both the central baryon and energy densities, as a larger matter pressure is necessary to balance the Lorentz force.  The addition of $\Delta$s decreases both quantities, as these stars are naturally (at B=0) smaller.

\begin{figure}
    \centering
    \includegraphics[angle=0,width=\linewidth]{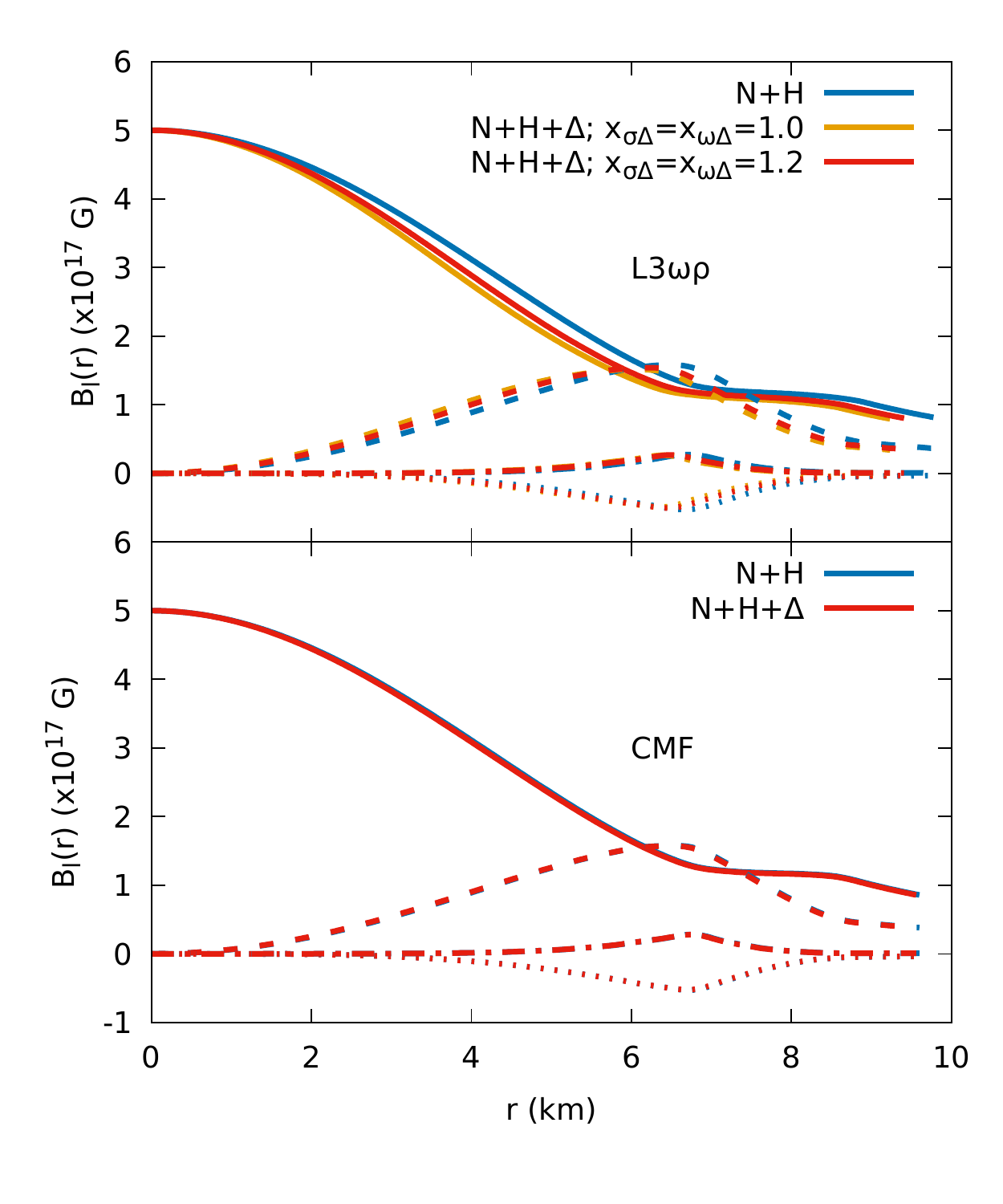}
    \vspace{-0.8cm}
    \caption{Magnetic field distribution inside a neutron star of mass $1.8M_{\odot}$ and central magnetic field of $B = 5\times 10^{17}$ G for different compositions and interaction strengths. Solid, dashed, dashed-dotted and dotted are, respectively, the first four even multipoles of the magnetic field norm ($l=0,2,4,6$), shown as functions of the coordinate radius. The top and bottom panels show results for the L3$\omega\rho$ and CMF models, respectively.
    }
    \label{fig:B_dist_1}
\end{figure}
 
\begin{figure}
    \centering
    \includegraphics[angle=0,width=\linewidth]{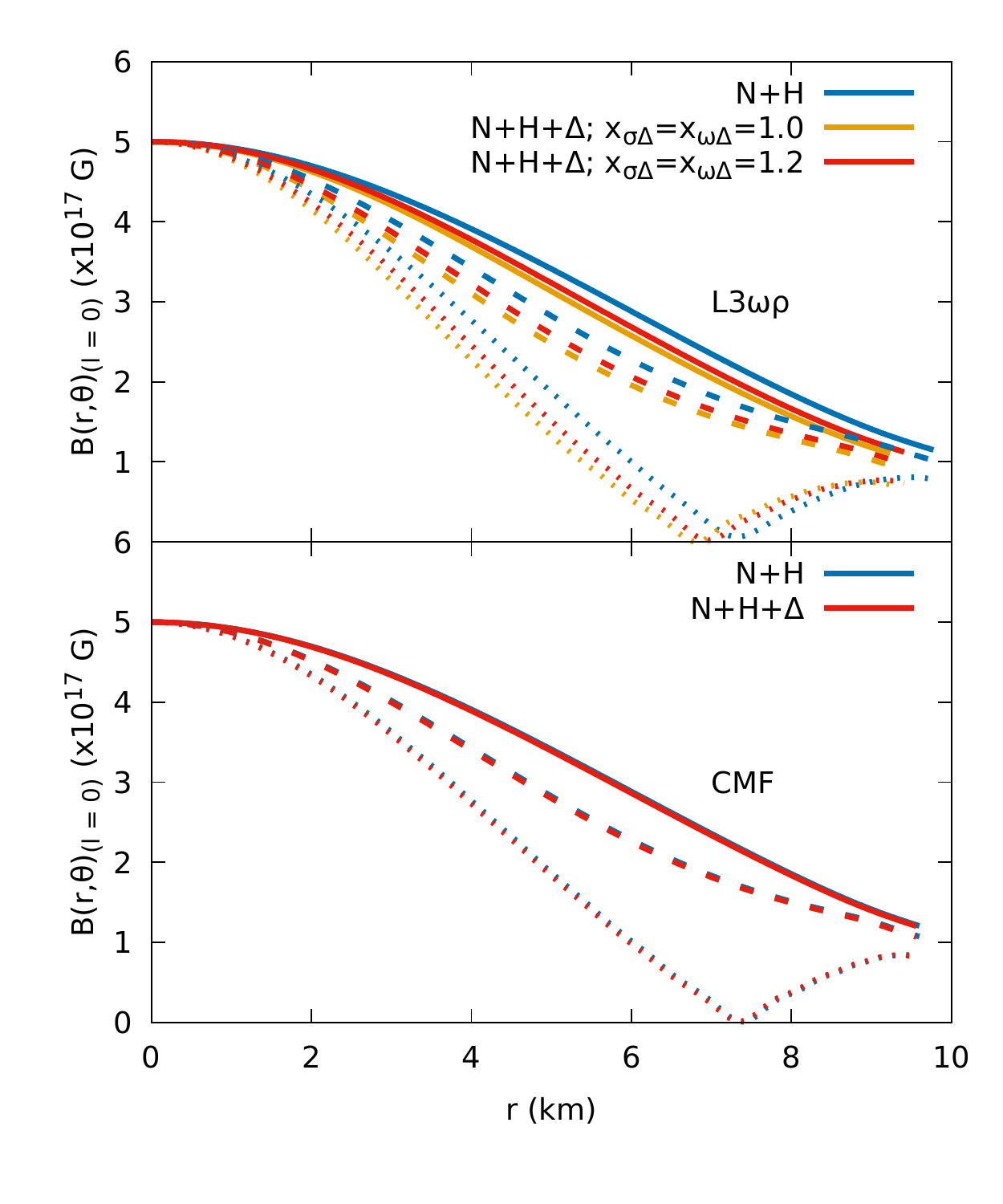}
    \vspace{-0.8cm}
    \caption{Magnetic field distribution inside a neutron star of mass $1.8M_{\odot}$ and central magnetic field of $B = 5\times 10^{17}$ G for different compositions and interaction strengths. Solid, dashed and dotted are the dominant monopolar ($l = 0$) term at the polar ($\theta=0$), intermediate ($\theta=\pi/4$) and equatorial ($\theta=\pi/2$) orientations, respectively, shown as functions of the coordinate radius. The top and bottom panels show results for the L3$\omega\rho$ and CMF models, respectively.
    }
    \label{fig:B_dist_2}
\end{figure}

At this point, we note that the maximum mass value of the stellar family described by the 
$L3\omega\rho$ model with $\Delta$s and $x_{\sigma \Delta} = 1.0$ (the yellow curve) has not been attained because, as explained earlier, the EoS numerical code stops converging at large densities due to reaching zero nucleon masses. Such behavior indicates that hadronic matter is no longer stable at this point and deconfinement to quark matter must be considered. We leave such analysis to a future work. But, since the trend of the yellow curve is quite obvious, we can conclude that its maximum mass is lower when compared to the other coupling and composition.

Using the full numerical solution, we also study the effect of the EoS on the magnetic field configurations inside a given star. We decompose the magnetic field norm in terms of spherical harmonics 
\begin{equation}\label{mag_decomp}
    B(r,\theta) \approx \sum_{i=0}^{l_{max}} B_l(r)Y_l^0(\theta)\ ,
\end{equation}
and plot the first four even multipoles ($l = 0,2,4,6$) as function of coordinate radius for both the EoS models and coupling strengths in Fig~\ref{fig:B_dist_1}. We also plot the profile of the dominant monopolar, spherically symmetric, term ($l = 0$) inside the star in Fig~\ref{fig:B_dist_2}. For $L3\omega\rho$ model, specially if we include $\Delta$s, the magnetic field norm decreases slightly inside the star but, for CMF model, we do not see any considerable changes.  

\begin{table}
    \centering
\caption{Central baryon ($n_c$) and energy ($\varepsilon_c$) densities as a function of magnetic field strength for neutron stars of radius 12 km with  L3$\omega\rho$ model for $x_{\sigma\Delta}=x_{\omega\Delta}=1.0 (1.2)$ in the top panel and CMF model in the bottom panel}
\label{tab:centralvalue}    
 \begin{tabular}{ccccc}
        \hline
        \multicolumn{1}{c}{\multirow{2}{*}{$B$ (G)}}   & \multicolumn{2}{c}{$n_c$ ($\rm{fm}^{-3}$)}  & \multicolumn{2}{c}{$\varepsilon_c$ ($\rm{MeV/fm^{3}}$)} \\ \cline{2-5}
        \multicolumn{1}{c}{}         \vspace{-.25cm}\\     \vspace{-.1cm}    
        & N+H & \begin{tabular}[c]{@{}c@{}}N+H+$\Delta$ \vspace{0.1cm}\\ \end{tabular}  &  N+H \vspace{0.1cm} &  \begin{tabular}[c]{@{}c@{}}N+H+$\Delta$ \vspace{0.1cm}\\ \end{tabular}                                  \\ \hline
        \multicolumn{1}{c}{0}          &  0.672  & 0.618 (0.614)   &  742  & 658 (657)  \\ 
        \multicolumn{1}{c}{$5\times 10^{17}$} & 0.701  & 0.659 (0.653)  & 783 & 712 (708)   \\ 
        \multicolumn{1}{c}{$1\times10^{18}$} & 0.747  & 0.714 (0.707)   & 850 & 786 (783)    \\ \hline
        
           \multicolumn{1}{c}{0}          &  0.629  & 0.625   &  678  & 672  \\ 
        \multicolumn{1}{c}{$5\times 10^{17}$} & 0.680  & 0.677  & 747 & 741   \\ 
        \multicolumn{1}{c}{$1\times10^{18}$} & 0.749  & 0.746   & 843 & 837    \\ \hline
        \vspace{-.3cm}
        \end{tabular}
\end{table}

\section{Conclusions}

In the present work, we have used two different relativistic models, namely, one of the parametrizations of the Walecka model with non-linear terms called  L3$\omega \rho$ and the chiral mean field (CMF) model, to investigate the effects of the presence of $\Delta$ baryons in dense matter. When the L3$\omega \rho$ is used, the unknown meson-hyperon coupling constants can be extracted from both phenomenology and symmetry group considerations, but the choice of meson-$\Delta$ couplings is still flexible. The CMF model, on the other hand, fixes the coupling constants with the help of phenomenological potentials uniquely. As a consequence, single particle $\Delta$ baryon potentials and the compressibility away from saturation are quantitatively different for both models, although they qualitatively present the same behavior. This is a consequence mainly of the different hyperon and $\Delta$ composition at intermediate and large densities. Those features are carried out when strong magnetic fields and anomalous magnetic moments (AMMs) are incorporated.

We carefully investigated particle composition and spin polarization when $\Delta$ baryons are included in neutron-star matter under the influence of strong magnetic fields with and without AMM corrections. Due to the effects of charge conservation and chemical equilibrium, there is no common behavior for all the particles (as predicted by their AMM signs and strengths). However, in general, while the population of charged particles increases with the inclusion of AMM, the population of neutral particles tends to decrease. 

The macroscopic properties of magnetars for the above choice of EoS models were obtained by solving Einstein-Maxwell equations within the LORENE library. It was found that maximum masses as high as 2$M_{\odot}$ can be attained even on inclusion of $\Delta$ particles. This is due to isospin readjustment at large densities, which turns the EoS stiffer. The $\Delta$s also respond more strongly to the AMM, which is expected due to the fact that they present additional electric charges and isospin projections. As a consequence, $\Delta$-admixed hypernuclear stellar matter, possesses larger spin polarization. The latter effect is more dramatic for the L3$\omega \rho$ model, which presents a larger number of exotic particles than the CMF model. 

Considering strong magnetic fields, heavy stars tend to contain more $\Delta$s in their interiors. They are not necessarily more massive (than their B=0 counterparts), but are larger and, for a given radius, present larger central number density and energy density. While $\Delta$s modify the magnetic field distribution very little inside stars, they decrease their radii, improving the agreement with modern observational data of neutron-star radii and tidal deformability \citep{Miller:2019cac,Riley:2019yda,Miller:2021qha,Riley:2021pdl,LIGOScientific:2018hze}.

Our results do not show the significant increase of stiffness of neutron-star matter with $\Delta$s at large density, as initially observed by \citet{Dexheimer:2021sxs}. In a more in depth analysis (to appear in a separate publication \citet{marquezyt}), we will show that this effect can indeed occur and reproduce more massive magnetars, but is very sensitive to model parameters.

\section*{Acknowledgements}

We thank Jorge Noronha for the very useful discussions.
This work is a part of the project INCT-FNA Proc. No. 464898/2014-5. K.D.M. acknowledges a doctorate scholarship from Conselho Nacional de Desenvolvimento Científico e Tecnológico (CNPq/Brazil). M.R.P. acknowledges a doctorate scholarship from Coordenação de Aperfeiçoamento de Pessoal do Ensino Superior (Capes/Brazil). D.P.M. was partially supported by Conselho Nacional de Desenvolvimento Científico e Tecnológico (CNPq/Brazil) under grant 303490-2021-7. V. D. acknowledges support from the National Science Foundation
under grants PHY1748621, MUSES OAC-2103680, and NP3M
PHY-2116686, in addition to PHAROS (COST Action CA16214).



\bibliography{references}

\end{document}